\documentclass[aps,prl,twocolumn,superscriptaddress]{revtex4}
\usepackage{amsmath,amssymb}
\usepackage{bm}
\usepackage{graphicx}
\usepackage{epstopdf}
\newcommand{\beq}{\begin{equation}}
\newcommand{\eeq}{\end{equation}}

\begin{document}
\title{Quasiparticle Pattern of Phenomena in Exotic Superconductors}
\author{V.~A.~Khodel}
\affiliation{National Research Centre Kurchatov
Institute, Moscow, 123182, Russia}
\affiliation{McDonnell Center for the Space Sciences \&
Department of Physics, Washington University,
St.~Louis, MO 63130, USA}
\author{J.~W.~Clark}
\affiliation{McDonnell Center for the Space Sciences \&
Department of Physics, Washington University,
St.~Louis, MO 63130, USA}
\affiliation{Centro de Investiga\c{c}\~{a}o em Matem\'{a}tica
e Aplica\c{c}\~{o}es, University of Madeira, 9020-105
Funchal, Madeira, Portugal}
\author{M.~V.~Zverev}
\affiliation{National Research Centre Kurchatov
Institute, Moscow, 123182, Russia}
\affiliation{Moscow Institute of Physics and Technology,
Dolgoprudny, Moscow District 141700, Russia}
	
\begin{abstract}

The quasiparticle formalism invented by Lev Landau for description
of conventional Fermi liquids is generalized to exotic superconductivity
attributed to Cooper pairing, whose measured properties defy explanation
within the standard BCS-Fermi Liquid description. We demonstrate that
in such systems the quasiparticle number remains equal to particle
number, just as in common Fermi liquids.  We are then able to explain
the puzzling relationship between the variation with doping $x$ of
two key properties of the family La$_{2-x}$Sr$_x$Cu0$_4$ of exotic
superconductors, namely the $T=0$ {\it superfluid} density
$\rho_{s0}(x)$ and the coefficient $A_1(x)$ in the linear-in-$T$
component of the {\it normal-state} low-$T$ resistivity
$\rho(T)=\rho_0+A_1T+A_2T^2$, in terms of the presence of
interaction-induced flat bands in the ground states of these metals.
\end{abstract}
\maketitle

The BCS paradigm~\cite{BCS,gor'kov,eliash}, emergent more than half a
century ago, has successfully explained the phenomenon of superconductivity
discovered by Kamerlingh Onnes in 1911. This success rests upon (i) the
Cooper scenario for electron pairing in metals~\cite{cooper} and (ii) the
Landau quasiparticle formalism, applicable to the normal state of a
Fermi liquid (FL) provided the damping $\gamma$ of single-particle
excitations is small compared with their energy $\epsilon({\bf p})$
measured from the chemical potential $\mu$~\cite{lan1,lan2}.  Subsequently,
Larkin and Migdal (LM) adapted the BCS-FL theory to quantitative
description of superfluid liquid $^3$He~\cite{LM,migdal}. One of the
prominent LM results is that the $T=0$ superfluid density $\rho_{s0}$
coincides with total density $\rho$, {\it irrespective} of the strength of
interparticle forces.

However, the LM theory fails to describe superconducting alloys.
In the presence of impurity-induced electron scattering, the damping
$\gamma$ becomes finite, rendering the Landau postulate
$\gamma/|\epsilon({\bf p})|\ll 1$ inapplicable. In the analysis
of the properties of these metals pioneered by Abrikosov and
Gor'kov(AG)~\cite{AG}, an additional dimensionless parameter
$\gamma/T_c(x)$ comes into play, resulting in substantial suppression
of $\rho_{s0}$ as observed experimentally, with $\rho_{s0}(x)$ {\it
coming  to naught} at a doping value $x_c$, in tandem with the critical
temperature $T_c(x)$.  Although the effects of $e-e$ interaction
are ignored in AG theory, their involvement within the
standard BCS-FL approach makes little difference~\cite{prb2019}. These
findings suggest that the replacement of FL quasiparticles by
more realistic quasiparticles of finite lifetime is instrumental
to elucidating the properties of superconducting alloys.

The BCS-FL-AG era ended dramatically with the discovery by Bednorz and
M\"uller (BM)~\cite{BM} of exotic superconductivity, whose properties
defy explanation within the BCS paradigm, opening up a new chapter of
condensed-matter physics devoted to studies of non-Fermi-liquid (NFL)
behavior of strongly correlated electron systems~\cite{leggett}.  Results
of extensive later experimental studies of the evolution of superfluid
density with doping $x$ and temperature $T$, performed in overdoped
high-$T_c$ superconducting LSCO compounds, have confirmed the
collapse of the BCS-FL-AG formalism~\cite{zaanen,bozovic,bozprl,bozltp}.
Given this situation, an implicit question drives our agenda:
Is it possible to further modify the Landau formalism so as to
adapt it to description of such NFL behavior, well documented in
recent years?  As will be seen, the answer to this question is
positive.

Any version of the quasiparticle pattern is based on decomposition of
the single-particle Green's function $G$ into the sum~\cite{lan2,AGD}
\beq
G({\bf p},\varepsilon)\equiv zG^q({\bf p},\varepsilon)+
G^r({\bf p},\varepsilon).
\label{dec}
\eeq
Here $G^r({\bf p},\varepsilon)$ is the regular part of $G$,
while $G^q({\bf p},\varepsilon)$, entering with quasiparticle weight $z$,
is the pole term.  In FL theory, one has
\beq
G^q({\bf p},\varepsilon)=\frac{1-n_L({\bf p})}{\varepsilon
-\epsilon({\bf p})+i\gamma(\varepsilon)} +\frac{n_L({\bf p})}
{\varepsilon-\epsilon({\bf p})-i\gamma(\varepsilon)}  ,
\label{lg}
\eeq
with the damping $\gamma$ small compared to $|\epsilon({\bf p})|$
and the Landau quasiparticle momentum distribution
\beq
n_L({\bf p})=\theta(-\epsilon({\bf p})),
\label{ldf}
\eeq
normalized by $\rho=(2/2\pi^3) \int n_L({\bf p})d^3{\bf p}$ .

FL theory is designed to express all low-$T$ characteristics of
Fermi systems in terms of the quasiparticle Green's functions
$G^q$ and a universal phenomenological interaction function
$f$ that absorbs all contributions from $G^r$.  An integral
feature of the FL pattern is equality between the particle
and quasiparticle numbers, known as the celebrated
Landau-L\"uttinger (LL) theorem.

In dealing with superconducting alloys, Eq.~(\ref{ldf}) still holds
when $\gamma$ becomes finite, while remaining small compared with the
bandwidth even in the dirtiest alloys.  Given the obvious violation
of the FL condition $\gamma/|\epsilon({\bf p})|\ll 1$, the
FL formalism has never been applied to check for any analogs of
the LL theorem  in these systems.  Furthermore, the authors of
some theoretical articles (see e.g.~\cite{pjh2}) claim that
violation of this condition rules out the possibility of creating
a quasiparticle pattern of phenomena in strongly correlated Fermi
systems. However, as we will see, this is not the case: the
quasiparticle picture can still apply, including the equality
between the quasiparticle and particle numbers, {\it at any
realistic value of the ratio $\gamma/|\epsilon({\bf p})|$}.

Upgrade of the FL proof of the LL theorem~\cite{AGD} is based on
analysis of specific behavior of a Fermi system placed in an external
long-wavelength longitudinal field ${\bf p}{\bf E}({\bf k},\omega)$.
While the effect of the field is absent in the limit ${\bf k}= 0,
\omega  \neq 0$, it becomes well pronounced in the opposite case
$\omega=0,{\bf k}\neq 0$, no matter how small the wave vector
${\bf k}$. To make proper use of this unique feature, we rewrite
the usual formula for $\rho$ in terms of the corresponding response
function:
$$
\rho=\frac {N}{V}= - \frac{2}{3}\int\!\!\!\!\int_C p_n
\frac{\partial G({\bf p},\varepsilon)} {\partial p_n}
\frac{d^3{\bf p}\,d\varepsilon}{(2\pi)^4i} \qquad\qquad\qquad
$$
\vskip -0.6 cm
\beq
=\frac{2}{3} \int\!\!\!\!\int_C\!\! p_n K({\bf p},\varepsilon)
\frac{\partial G^{-1}({\bf p},\varepsilon)}
{\partial p_n} \frac{ d^3{\bf p}\,d\varepsilon}{(2\pi)^4i} ,
\label{pn}
\eeq
where $p_n$ is the normal component of momentum
${\bf p}$ and $K({\bf p},\varepsilon)= \lim_{{\bf k}\to 0}
G({\bf p},\varepsilon) \,G({\bf p}{+}{\bf k},\varepsilon)$.
That the integral~(\ref{pn}) does represent the longitudinal
response function follows from the relation
${\cal T}({\bf p},\varepsilon; {\bf k}\to 0,\omega=0)
=-\partial G^{-1}({\bf p},\varepsilon)/ \partial {\bf p}$
based on gauge invariance~\cite{AGD}.

In accord with results from pioneering work of Migdal~\cite{migp},
$\rho$ decomposes into a sum $\rho=\rho_L+\rho_R$, with
\begin{eqnarray}
\rho_L&=&
\frac{2}{3} \int\!\!\!\!\int_L\!\! p_n K({\bf p},\varepsilon)
\frac{\partial G^{-1}({\bf p},\varepsilon)}
{\partial p_n} \frac{ d^3{\bf p}\,d\varepsilon}{(2\pi)^4i}, \nonumber\\
\rho_R&=&
\frac{2}{3} \int\!\!\!\!\int_R\!\! p_n G({\bf p},\varepsilon)
\,G({\bf p},\varepsilon) \frac{\partial G^{-1}({\bf p},\varepsilon)}
{\partial p_n} \frac{ d^3{\bf p}\,d\varepsilon}{(2\pi)^4i} .
\label{pnk}
\end{eqnarray}
\begin{figure}[t]
\begin{center}
\includegraphics[width=1.\linewidth] {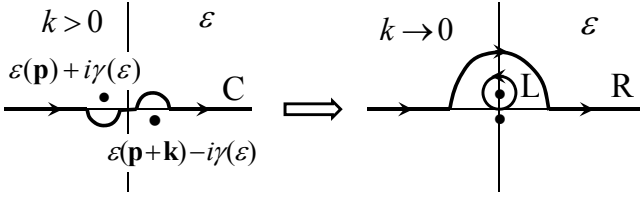}
\end{center}
\vskip -0.5 cm
\caption{Arrangement of the contour $C$ for the energy integration in
Eq.~(\ref{pn})}
\label{fig:LL}
\end{figure}

The term $\rho_L$ containing a loop integral absorbs quasiparticle
contributions from the poles of $G^q$ having the form~(\ref{lg}).
Implicitly, quasiparticle contributions are also present in a
term $\rho_R$ associated with integration along the remaining part
$R$ of the contour $C$ (see Fig.~1). To prove this we employ the
relation~\cite{AGD}
\beq
- \nabla G^{-1}({\bf p},\varepsilon)
= {\bf v}_0({\bf p}) + \int\!\!\!\!\int_C\!\!
{\cal U}({\bf p},\varepsilon,{\bf l},\omega) \nabla G({\bf l},\omega)
\frac{d^3{\bf l}\,d\omega}{(2\pi)^4i}
\label{pit60}
\eeq
derived within many-body theory assuming gauge invariance.  Here
${\bf v}_0=\nabla\epsilon^0_{{\bf p}}$ is the bare group velocity,
while ${\cal U}$ represents the block of Feynman diagrams for the
scattering amplitude irreducible in the particle-hole channel, and
$\nabla=\partial/\partial{\bf p}$.

The first step of our program, adapted from FL theory, is implemented
by introducing an interaction amplitude $\Gamma^R$ determined by the
Landau equation~\cite{lan2,AGD}
$$
\Gamma^R({\bf p},\varepsilon,{\bf p}_1,\varepsilon_1)={\cal U}({\bf p},
\varepsilon,{\bf p}_1,\varepsilon_1)
$$
\vskip -0.8 cm
\beq
+2 \int\!\!\!\!\int_R\!\!
{\cal U}({\bf p},\varepsilon,{\bf l},\omega)G({\bf l},\omega)G({\bf l},
\omega) \Gamma^R({\bf l},\omega,{\bf p}_1,\varepsilon_1)
\frac{d^3{\bf l}\,d\omega} {(2\pi)^4i}  .
\label{gammr}
\eeq
Hereafter we employ FL symbolic notations, with round brackets implying
summation and integration over all intermediate variables, supplemented
by respective normalization factors. Thereby Eq.~(\ref{gammr}) becomes
\beq
\Gamma^R={\cal U}+\Bigl(\Gamma^R K{\cal U}\Bigr)_{\!R}\equiv {\cal U}
+\Bigl({\cal U}K\Gamma^R\Bigr)_{\!R} .
\label{gamr}
\eeq
Further, as usual, we multiply Eq.~(\ref{pit60})
from the left by $\Gamma^RGG$ and perform R-integration to obtain
\beq
- \nabla G^{-1}={\bf v}_0+\Bigl(\Gamma^R K {\bf v}_ 0\Bigr)_{\!R}
+\Bigl(\Gamma^R \nabla  G\Bigr)_{\!L} .
\label{ptr}
\eeq
Both Eqs.~(\ref{gamr}) and (\ref{pit60}) were employed to obtain
this result.  Upon its substitution into the second integral of
Eq.~(\ref{pnk}), one finds
$$
\rho_R=\Bigl({\bf p} K\nabla G^{-1}\Bigr)_{\!R}=-\biggl(\Bigl[{\bf p}
+\Bigl({\bf p}K\Gamma^R\Bigr)_{\!R}\Bigr]  K  {\bf v}_0
\biggr)_{\!R}
$$
\vskip -0.5 cm
\beq
+\biggl(\!\Bigl({\bf p}K\Gamma^R\Bigr)_{\!R} K\nabla G^{-1}\biggl)_{\!\!L}.
\label{reli}
\eeq
After employing the relation~\cite{migdal,pit}
\beq
\frac{\partial G^{-1}({\bf p},\varepsilon)}{\partial\varepsilon}{\bf p}
=  {\bf p}+ \Bigl({\bf p}K\Gamma^R\Bigr)_{\!R},
\label{pm}
\eeq
applicable provided the momentum operator ${\bf p}$ commutes with the
total Hamiltonian of the system, Eq.~(\ref{reli}) is significantly
facilitated, taking the form
\beq
\rho_R = \Bigl({\bf p}\Bigl[\frac{\partial G^{-1}({\bf p},\varepsilon)}
{\partial\varepsilon}-1\Bigr] K\nabla G^{-1}\Bigr)_{\!\!L} ,
\label{rhor}
\eeq
upon noting that the first term on the r.h.s. of Eq.~(\ref{reli}),
rewritten as $-\Bigl(p_n(\partial G^{-1}/\partial\varepsilon)GGv^0_n\Bigr)_{\!R} = \Bigl(p_n v^0_n \partial G/\partial\varepsilon\Bigr)_{\!R}$, vanishes
upon energy integration.

Summation of $\rho_R$ from Eq.~(\ref{rhor}) and $\rho_L$ as given by
the first of the integrals~(\ref{pnk}) yields the desired result
\beq
\rho= \frac{2}{3}\int\!\!\!\!\int_L\!\!\frac{\partial G^{-1}({\bf p},
	\varepsilon)}{\partial\varepsilon } p_n  K({\bf p},\varepsilon)
\frac{\partial G^{-1}({\bf p},\varepsilon)} {\partial p_n}
\frac{ d^3{\bf p}\,d\varepsilon}{(2\pi)^4i} .
\label{pnq}
\eeq
Near the pole, one has $\partial G^{-1}({\bf p},
\varepsilon)/\partial\varepsilon =z^{-1}$, while
$\nabla G^{-1}({\bf p}, \varepsilon)=- z^{-1}\nabla\epsilon({\bf p})$.
Given that the Fermi surface (FS) remains simply connected, insertion
of these results into Eq.~(\ref{pnq}) produces
\beq
\rho =
-\frac{2}{3}\int\!\!\!\!\int_L p_n\frac{\partial G^q({\bf p},\varepsilon)}
{\partial p_n} \frac{d^3{\bf p}\,d\varepsilon}{(2\pi)^4i}
=2\!\int n_L({\bf p})\frac{d^3p}{(2\pi)^3}=\frac{p^3_F}{3\pi^2}.
\label{llg}
\eeq
This result, known as the Landau-L\"uttinger (LL) theorem, remains valid
as long as the equation
\beq
\epsilon({\bf p},n_L)=0
\label{root}
\eeq
has a single root~\cite{lifshitz,volovik}. This is indeed the case,
provided the change $\delta E(n_L)=\sum_{{\bf p}}
\epsilon({\bf p};n_L)\delta n_L({\bf p})$ of the energy of the
Landau state remains non-negative under {\it any variation} of the
momentum distribution $n_L({\bf p})$ compatible with the Pauli
principle~\cite{physrep}. This is true for homogeneous Fermi liquids
where $\epsilon(p,n_L)=v_F(\rho)(p-p_F)$, provided the Fermi velocity
$v_F=p_F/m^*$ remains positive~\cite{lan1}. It then follows that
the quantities $\epsilon({\bf p})$ and $\delta n({\bf p})$ always
have the same sign, to guarantee $\delta E>0$.

Analogous manipulations performed for Eq.~(\ref{pit60}) lead to the
Pitaevskii equation~\cite{pit}
\beq
\frac{\partial\epsilon({\bf p};n_L)}{\partial {\bf p}}=
\frac{\partial\epsilon_0({\bf p})}{\partial {\bf p}}
+2\int f({\bf p},{\bf l})\frac{\partial n_L({\bf l})}
{\partial {\bf l}} \frac {d^3 {\bf l}}{(2\pi)^3}
\label{pitfl}
\eeq
involving the interaction function $f=z^2\Gamma^R$. Given its form,
Eq.~(\ref{pitfl}) can be solved numerically to yield the quasiparticle
spectrum $\epsilon({\bf p};n_L)$ in all of momentum space
\cite{physrep,zkp}.  However, Eqs.~(\ref{llg}) and (\ref{pitfl}) need
to be rearranged when Eq.~(\ref{root}) acquires additional roots
that occur if the Fermi velocity $v_F$, calculated for the given
Landau state, changes sign.  In the 2D homogeneous electron liquid
of MOSFETs, such a situation occurs at a critical density
$\rho_\infty=0.8\times 10^{11}\,{\rm cm}^{-2}$ \cite{mokashi} where
both the density of states and the effective mass diverge. Beyond
this topological critical point (TCP), {\it countless} options
for breakdown of the original Landau state arise.

The anisotropy of the electron spectrum in solids furnishes additional
opportunities for topological rearrangement of the Landau state.  These
effects are associated with the inflow of the TCPs where the function
$v_F(p,{\bf  n};n_L)$ found from Eq.~(\ref{pitfl}) changes sign at
certain points of the Fermi surface, occurring automatically if the
respective solutions of Eq.~(\ref{pitfl}) attain boundaries of the
Brillouin zone.  Presumably, such a situation is realized in twisted
bilayer graphene (TBLG), where $v_F(\rho,\theta)$ passes through zero
at a critical twist angle $\theta_m\simeq 1.1^{\circ}$, inducing
an {\it inevitable topological rearrangement} of nearly-flat-band
solutions, which have been identified in Ref.~\cite{bm}. In this
case, variations $\delta E(n_L)$ inescapably acquire a negative sign
ubiquitously in the whole momentum region where $v_F(n_L)<0$, implying
that the number of roots of Eq.~(\ref{root}) becomes {\it infinite} again.

A relevant solution of the problem can be found, requiring the
associated energy variations
\beq
\delta E(n_*)=\sum_{\bf p} \epsilon({\bf p};n_*)\delta n_*({\bf p})
\label{scfc}
\eeq
of the state with the rearranged quasiparticle momentum distribution
$n_*(p)$ to be non-negative.  Allowing the permissible occupation
numbers $n({\bf p};n_*)$ to lie between 0 and 1, both signs of
$\delta n_*({\bf p})$ come into play.  Non-negativity of $\delta E(n_*)$
can then be ensured, provided the energy $\epsilon({\bf p},n_*)$
vanishes {\it identically} in the momentum region $\Omega$. Accordingly,
in this regime the single-particle spectrum becomes {\it completely
flat}~\cite{ks,vol1, noz,ktsn1,volovik,prb2008,m100,annals,ktsn2,book}
yielding
\beq
0=\frac{\partial\epsilon_0({\bf p})}{\partial {\bf p}}+2\int
f({\bf p},{\bf l})\frac{\partial n_*({\bf l})}{\partial {\bf l}}
\frac {d^3 {\bf l}}{(2\pi)^3} , \quad {\bf p}\in \Omega,
\label{eqfc}
\eeq
while remaining unchanged outside $\Omega$ (except for the obvious
replacement $n_L\to n_*$).

Previously \cite{prb2008,m100}, we have investigated the fate of
the LL theorem in Fermi systems harboring the fermion condensate (FC),
where the pole term $G^q$ becomes
\beq
G^q({\bf p},\varepsilon)=\frac{1-n_*({\bf p})}{\varepsilon
-\epsilon({\bf p}) +i\gamma(\varepsilon)} +\frac{n_*({\bf p})}
{\varepsilon-\epsilon({\bf p}) -i\gamma(\varepsilon)} ,
\label{polfc}
\eeq
with $\epsilon({\bf p})$ now determined from Eq.~(\ref{eqfc}).
In Refs.~\cite{prb2008,m100}, we have obtained the relation
\beq
\rho=2\int n_*({\bf p})\frac{d^3{\bf p}} {(2\pi)^3},
\label{llfc}
\eeq
which also follows from Eq.~(\ref{llg}) upon inserting Eq.~(\ref{polfc}).

A salient feature inherent in states having an interaction-induced
flat band is exhibited in the advent of an entropy excess $S_*$ given
by Landau-like formula
\beq
\frac{S_*}{V}=-2\int_\Omega[ n_*({\bf p})\ln n_*({\bf p})
+ (1-n_*({\bf p}))\ln (1-n_*({\bf p})] \frac{d^3{\bf p}}{(2\pi)^3} .
\label{se}
\eeq
In essence, Eqs.~(\ref{eqfc})-(\ref{se}) form the basis of the
interaction-induced flat-band scenario, also called the theory of
fermion condensation.

Adaptation of the foregoing strategy to the description of
superconducting alloys naturally requires the introduction of
Gor'kov equations involving two different single-particle Green's
functions~\cite{AGD,gor'kov,migdal,gor},
\begin{eqnarray}	
G_s({\bf p},\varepsilon)&=&\Bigl[G^{-1}({\bf p},\varepsilon)
+\Delta^2({\bf p})G(-{\bf p},-\varepsilon)\Bigr]^{-1} , \nonumber \\
F({\bf p},\varepsilon)&=&G(-{\bf p},-\varepsilon)\Delta({\bf p})
G_s({\bf p},\varepsilon) .
\label{gorf}
\end{eqnarray}
Here the normal-state Green's function $G({\bf p},\varepsilon)$ obeys
formulas~(\ref{dec}) and (\ref{lg}), as before.

Within the framework of the BCS approach, the superconducting
gap $\Delta$ is supposed to be $p-$ and $\varepsilon-$independent,
which greatly facilitates further analysis. Eq.~(\ref{pn}) is
then replaced by
$$
\rho= - \frac{2}{3}\int\!\!\!\!\int_C p_n\frac{\partial G_s({\bf p},
\varepsilon)} {\partial p_n} \frac{d^3{\bf p}\,d\varepsilon}{(2\pi)^4i}$$
\beq
=\frac{2}{3} \int\!\!\!\!\int_C\!\! {\bf p}K_s({\bf p},\varepsilon) \nabla
G^{-1}({\bf p},\varepsilon) \frac{ d^3{\bf p}\,d\varepsilon}{(2\pi)^4i} ,
\label{rhos}
\eeq
where
\beq
K_s({\bf p},\varepsilon)=\lim_{{\bf k}\to 0} [G_s({\bf p},\varepsilon)
G_s({\bf p} +{\bf k},\varepsilon)-F({\bf p},\varepsilon)
F({\bf p}+{\bf k},\varepsilon)] .
\label{kss}
\eeq
In symbolic notations, one now has
\begin{eqnarray}
\rho_R&=&\Bigl({\bf p} K_s\nabla G^{-1}\Bigr)_{\!R} , \quad
\Gamma^R={\cal U}+\Bigl(\Gamma^R K_s{\cal U}\Bigr)_{\!R} , \nonumber\\
- \nabla G^{-1}&=&{\bf v}_0+\Bigl(\Gamma^R K_s {\bf v}_ 0\Bigr)_{\!R}
+\Bigl(\Gamma^R \nabla  G\Bigr)_{\!L} ,\nonumber\\
\frac{\partial G^{-1}({\bf p},\varepsilon)}{\partial\varepsilon}{\bf p}
&=&  {\bf p}+ \Bigl({\bf p}K_s\Gamma^R\Bigr)_{\!R}.
\label{pits}
\end{eqnarray}
These formulas are obtained from those derived for a normal Fermi
liquid through the replacement $K\to K_s$.

Proceeding farther along the same lines as before, we find
\beq
\rho_R=-\Bigl(\frac{\partial G^{-1}}{\partial\varepsilon}K_s{\bf v}_0
\Bigr)_{\!R}+
\Bigl({\bf p}\Bigl[\frac{\partial G^{-1}({\bf p},\varepsilon)}
{\partial\varepsilon}-1\Bigr] K_s\nabla G^{-1}\Bigr)_{\!\!L} .
\label{relis0}
\eeq
The first term in the sum vanishes again, since $K_s\partial
G^{-1}/\partial\varepsilon\equiv \partial G_s/\partial \varepsilon$,
and hence its integration over energy
vanishes.  Thus we arrive at a nontrivial result: regular (R)
contributions to the density $\rho$ associated with the contour
R that may in principle depend on the gap value drop out
{\it identically}, so we are left with the pole contributions (L)
tied to the loop contour L. Indeed, upon summation of $\rho_R$
with $\rho_L$, we are led to
\beq
\rho= \frac{2}{3}\int\!\!\!\!\int_L\!\!\frac{\partial G^{-1}({\bf p},
	\varepsilon)}{\partial\varepsilon } p_n  K_s({\bf p},\varepsilon)
\frac{\partial G^{-1}({\bf p},\varepsilon)} {\partial p_n}
\frac{ d^3{\bf p}\,d\varepsilon}{(2\pi)^4i} .
\label{rhosq}
\eeq
Near the quasiparticle pole $ G_s({\bf p},\varepsilon)
=zG^q_s({\bf p},\varepsilon)$ and $ K_s({\bf p},\varepsilon)=z^2
[G_s^q({\bf p},\varepsilon)G_s^q({\bf p},\varepsilon)
-F^q({\bf p},\varepsilon)F^q({\bf p},\varepsilon)]$,
with~\cite{LM,migdal}
\beq
G_s^q({\bf p},\varepsilon)=\frac{u^2({\bf p})}{\varepsilon-E({\bf p})
+i\delta}+ \frac {v^2({\bf p})}{\epsilon +E({\bf p})-i\delta}  ,
\label{gsq}
\eeq
and $v^2({\bf p})=(E({\bf p})-\epsilon({\bf p}))/2E({\bf p})$, where
$E({\bf p})=\sqrt{\epsilon^2({\bf p})
+\Delta^2({\bf p})}$ is the Bogolyubov quasiparticle energy.  Upon performing
loop integrations in Eq.~(\ref{rhosq}), all the $z$-factors cancel
out again to arrive at
\beq
\rho= - \frac{2}{3}\int\!\!\!\!\int p_n
\frac{\partial G_s^q({\bf p},\varepsilon)} {\partial p_n}
\frac{d^3{\bf p}\,d\varepsilon}{(2\pi)^4i}
=2\!\int \! v^2({\bf p})\frac{d^3{\bf p}}{(2\pi)^3} .
\label{lls}
\eeq
Therewith we have demonstrated the coincidence between the particle and
quasiparticle densities in Cooper superconductors, irrespective of
the ratio $\Delta/T_F$ and the magnitude of the damping $\gamma$ in
normal states.

We are now in a position to analyze one of the most challenging results
of recent extensive experimental studies of overdoped LSCO compounds.
This is the deep connection between anomalous properties of their
superconducting and normal states, revealed by comparison of the critical
temperature $T_c(x)$ of termination of exotic superconductivity with
the linear-in-$T$ term $A_1(x)$ in the low-$T$ normal-state resistivity
$\rho(T>T_c)=\rho_0+A_1T+A_2T^2$ (identified over a decade ago in
Refs.~\cite{hussey1,paglione}).  This connection is exhibited in a
striking correlation between variations of the $T=0$ LSCO {\it superfluid
density} $\rho_{s0}(x)$ with doping $x$~\cite{zaanen} and the {\it
normal-state} coefficient $A_1(x)$~\cite{bozovic}.  Permanence of the
ratio ${\cal R}(x)=A_1(x)/\rho_{s0}(x)$ as a function of doping $x$,
as confirmed by data shown in Fig.~2, rules out all attempts to explain
the outstanding experimental results of Refs.~\cite{bozovic,bozltp,bozprl}
within the BCS-AG concept and its modifications. This includes the
scaling theory of Refs.~\cite{broun1,broun2}, where the $e-e$
interactions are completely ignored.  There the theoretical value
of the ratio ${\cal R}(x)$ is {\it identically} zero, since the
NFL effects are not accounted for in the BCS-AG approach, and
therefore $A_1$ is simply nonexistent.

\begin{figure}[t]
\begin{center}
\includegraphics[width=0.8\linewidth] {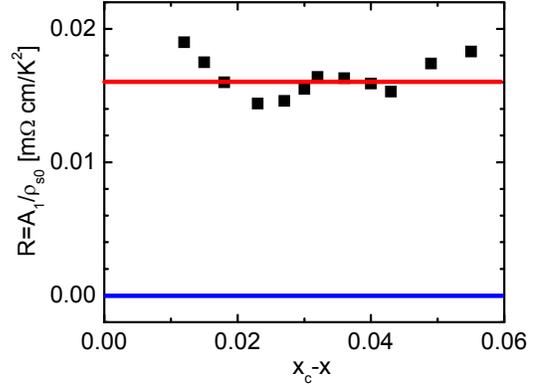}
\end{center}
\vskip -0.5 cm
\caption{Ratio ${\cal R}(x)=A_1(x)/\rho_{s0}(x)$ of the coefficient $A_1$ of
the linear-in-$T$ term in the low-$T$ normal-state resistivity of
La$_{2-x}$Sr$_x$CuO$_4$ compounds to their $T=0$ superfluid density
$\rho_{s0}$, versus doping $x$ measured from its critical value $x_c$
for gap termination.  Black squares show ${\cal R}(x)$ extracted from
the data of Ref.~\cite{bozovic}. The horizontal red line illustrates
the prediction for ${\cal R}(x)={\rm const.}$ within the FC scenario,
its value being chosen to match the experimental data, while the blue
line shows the zero value of this ratio within the BCS-AG concept.}
\label{fig:ratio}
\end{figure}

On the other hand, the experimental behavior of
$A_1(x)\propto x_c-x$~\cite{hussey1,paglione,bozovic}
is properly explained within the FC scenario, where its value
\beq
A_1(x)\propto \rho_{FC}(x)
\label{aifc}
\eeq
turns out to be proportional to the density $\rho_{FC}$ of the fermion
condensate. (For details, we refer the reader to recent
articles~\cite{jetplett2015,PLA2018,RC2020}).

Evaluation of the superfluid density $\rho_s$ reduces to finding the
response function $Q_{ij}$ that connects a $T=0$ electric current
${\bf j}$ with the {\it transverse} vector potential ${\bf A}$~\cite{AGD},
\beq
j_i({\bf k}) =-\frac {\rho e^2}{m_e} Q_{ij}({\bf k}) A_j({\bf k}).
\label{pen}
\eeq
One has $Q_{ij}(k)=(\delta_{ij}-k_ik_j/k^2)Q(k)$ and $\rho_{s0} =
Q(0)\rho $.  The function $Q(0)$ is known to contain a vacuum
contribution $Q_{\rm vac}=1$ coming from the term $\delta^2{\cal H}
=e^2A^2/2m_e$ in the corresponding second variation of the vacuum
Hamiltonian ${\cal H}$, which responsible, notably, for light
scattering by electrons. Thereby one obtains $Q(0)=1+P_s(0)$.
Importantly, in evaluation of the current-current correlator $P_s(0)$,
the propagator $L_s=G_sG_s+FF$ replaces $K_s=G_sG_s-FF$~\cite{AGD}
(which enters in the above proof of the LL-like theorem in
superconducting systems). Otherwise, the renormalization is carried
out along the same lines as in the foregoing proof of the equality
between the quasiparticle and particle numbers to yield~\cite{prb2019}
\beq
\rho_{s0}(x)/\rho=I(x)/[1+(\alpha-1)(1-I(x))] ,
\eeq
where $\alpha=m^*/m_e$ and
\beq
I(x)\propto\int\!\!\!\int F^2(\epsilon,\zeta)\,d\epsilon\,
d\zeta=\int\!\!\!\int\frac{\Delta^2 \eta^2(\zeta)\,d\epsilon\,d\zeta}
{\Bigl((\zeta^2+\Delta^2)\eta^2(\zeta)+\epsilon^2\Bigr)^2},
\eeq
with
\vskip -0.3 cm
$$
\eta(\zeta)=\frac{1}{2}\left(1+\frac{\gamma}{2\sqrt{\zeta^2
+\Delta^2}}\right)
\qquad\qquad\qquad\qquad
$$
\vskip -0.5 cm
\beq
+\left[\frac{1}{4}\left(1+\frac{\gamma}{2\sqrt{\zeta^2+\Delta^2}}
\right)^2 + \frac{\gamma}{\pi}\frac{\rho_{FC}}{\rho}\frac{\varepsilon_F}
{\zeta^2+\Delta^2}\right]^{1/2}.
\label{int}
\eeq
The FC contribution to the
integral (\ref{int}) is found to be insignificant at small FC density
$\rho_{FC}$ because this contribution is proportional to
$\Delta_0\rho_{FC}$.  Thus, the result of our calculations
\cite{prb2019}, namely
\beq
\rho_{s0}(x)\propto \Delta_0(x)\frac{m_e}{\gamma_{tr}m^*},
\eeq
turns out to be correct at any $x_c-x$.  Since the gap value
$\Delta_0$ is proportional to the FC density $\rho_{FC}$ as well \cite{ks,jetpl2017},
the function ${\cal R}(x) $ is indeed doping-independent, in
agreement with experiment.

This article is a logical complement to earlier work addressing the origin
of topological disorder \cite{RC2020} arising in strongly correlated
electron systems.  The quasiparticle formalism developed here furnishes
the proper theoretical foundation for the analysis of such phenomena.
Importantly, this formalism applies to superconducting
states with nontrivial topology as well, providing the basis for
{\it quantitative} analysis of interaction-induced effects in cuprates
and other high-$T_c$ superconductors, including magic-angle TBLG
where the standard near-flat-band solutions~\cite{bm} must experience
a topological rearrangement.

In conclusion, the authors are deeply grateful to V.\ Shaginyan
and G.\ Volovik for discussing issues addressed in this article.

\end{document}